\documentclass{article}

\PassOptionsToPackage{numbers, compress}{natbib}

\usepackage[final]{neurips_2021}

\usepackage[utf8]{inputenc} % allow utf-8 input
\usepackage[T1]{fontenc}    % use 8-bit T1 fonts
\usepackage{hyperref}       % hyperlinks
\usepackage{url}            % simple URL typesetting
\usepackage{booktabs}       % professional-quality tables
\usepackage{amsfonts}       % blackboard math symbols
\usepackage{nicefrac}       % compact symbols for 1/2, etc.
\usepackage{microtype}      % microtypography
\usepackage{xcolor}         % colors
\usepackage{graphicx}
\usepackage{graphics}
\usepackage{amsmath}
\usepackage{wrapfig}

\title{Towards Conditional Generation of Minimal Action Potential Pathways for Molecular Dynamics}
\author{%
  John Kevin Cava \\
  \And 
  John Vant \\
  \And
  Nicholas Ho \\
  \And
  Ankita Shukla \\
  \And
  Pavan Turaga \\
  \And
  Ross Maciejewski \\
  \And
  Abhishek Singharoy \\
  \And \\
  Arizona State University \\
  \{jcava,jvant, nichola2, ashukl20, pturaga, rmacieje, asinghar\}@asu.edu
}

\begin{document}

\maketitle

\begin{abstract}
 In this paper, we utilized generative models, and reformulate it for problems in molecular dynamics (MD) simulation, by introducing an MD potential energy component to our generative model. By incorporating potential energy as calculated from TorchMD into a conditional generative framework, we attempt to construct a low-potential energy route of transformation between the helix~$\rightarrow$~coil structures of a protein. We show how to add an additional loss function to conditional generative models, motivated by potential energy of molecular configurations, and also present an optimization technique for such an augmented loss function. Our results show the benefit of this additional loss term on synthesizing realistic molecular trajectories.

\end{abstract}

\section{Introduction}

% JV Intro
The structure and dynamics of proteins are central to our knowledge of molecular biology. 
Protein conformational changes are intimately connected to biological function. For example, some motor proteins changes their local shape from a helix to a coil-like form to force catalytic sites to change conformations in binding reacting and releasing sequences. Complementing our knowledge of 3-dimensional atomic structures, molecular dynamics or MD simulations has allowed researchers to investigate protein function by modeling such conformational changes.  The two most important parameters for tracking a conformational change are: a) the transition rate between shapes, and b) each shape's relative equilibrium population. Determining these kinetic and thermodynamic parameters  with MD enables a mechanistic description of protein functions.  

Large-scale conformational rearrangements involve non-linear and concerted movement of many atoms. Recent research in MD simulations has resulted in many advanced sampling techniques to extract kinetic and thermodynamic parameters from molecular data. However, in order to extract useful dynamics information, knowledge of the minimum-action or low-energy pathways is essential. The energetic barriers crossed by a non-optimal path will cause an erroneous estimation of the transition rate between states. Some notable examples of transition  path finding algorithms include anisotropic network models \cite{atilgan2001anisotropy}, the steepest decent dynamics \cite{chen2013efficiently}, steered MD~\cite{izrailev1999steered,park2004calculating} and the string simulations \cite{pan2008finding}.
The two main disadvantages to current path optimization algorithms are: (i) the computational expense, and, (ii) the complexity and skill required to set up the calculations.
Here, we present a pathway optimization algorithm to derive the minimum energy pathway between two protein conformational states by starting from a set of trial pathways.  

We utilize a conditional GAN architecture, in which the time-step of molecular dynamics is used as a conditional for generation. The generator creates a protein model of vector-size $3N$ for an $N$-atomic system, each atom  described with a feature-size of 3.
The potential energy along the backbone angles of these models are computed using the TorchMD package to constrain the generation, and also as an input for the discriminator. In order to increase quality of the model, we performed a pre-training of the generator with minimization on the dihedral angles and potential energy.

The stretching of deca-alanine from a coiled alpha-helix to an uncoiled stretched conformational state is used as a model system for protein conformational changes, due to its small size and relatively simple dynamics. The generated pathways were compared to ones derived from string with steered MD simulations and were in good agreement. The github link to the code and data: \url{https://github.com/johncava/Molecular_Dyanmics_cGAN}

\section{Related Work}

A number of approaches are available for learning and generation physical trajectories using deep learning models. Many previous works that use deep learning to model physical systems rely heavily on using a model to approximate a differential equation. Neural ordinary differential equations use the adjoint sensitivity method to update neural network parameters to approximate a continuous differential equation \cite{chen2018neural}. Hamiltonian Neural Networks (HNNs) predict a scalar value and take the symplectic gradient in order to approximate a differential equation that conserves total energy \cite{greydanus2019hamiltonian}. Symplectic-RNN attempts to improve on the stability of HNNs by introducing recurrent training \cite{chen2019symplectic}. Symplectic-ODE Net is a different approach that uses multiple neural networks to approximate a scalar value to which they take the symplectic gradient and use adjoint sensitivity to update the parameters \cite{zhong2019symplectic}. All of these models are useful for modeling non-noisy deterministic few-body systems, however, MD simulations are inherently stochastic many-body systems due to thermal fluctuations that follow the Langevin equations. The slow changes in reaction coordinate (such as collective angle or distances) along a pathway takes place over a much longer period of time than that of the atomistic thermal fluctuations. Another approach is to use a deep model to generate a new frame given a set of history frames using, by using temporal dynamical models such as LSTMs \cite{tsai2020learning,gupta2020mind}. Other works have used Langevin dynamics to find stable conformations \cite{shi2021learning}.
Lastly, methods have used HNNs with GNNs in order to train on data to enforce a potential energy bias in order to propel a string of beads in an MD simulation towards a desired shape (by demonstrating differentiable control) \cite{wang2020differentiable}. However, the determination of the potential energy from the data may not provide a pathway, or even if it is guaranteed to be minimal. In addition, we want to generalize the model to predict outside the stable conformation, such that we can hopefully generate non-equilibrium trajectories.

\section{Preliminaries}

\subsection{Analytical Potentials}\label{Sec:EnergyTerm}

In this paper, we utilize TorchMD \cite{doerr2020torchmd} to calculate the potential energy terms of a structure at a given MD time-step. We then use these potentials in conjunction with cGANs for the generator to learn realistic, low-energy structural models.
%Next, we enumerate the potential energy terms that TorchMD calculates, so we can further discuss them in the following sections.

% \begin{tabular}{ |p{3cm}||p{9cm}||  }
%  \hline
%  \multicolumn{2}{|c|}{Table 1: Analytical Potentials} \\
%  \hline
%  Bonds   & \begin{equation}
% V_{bonded} = k_0 (r - r_{eq})^2 
% \end{equation} \\
%  \hline
%  Angles &   \begin{equation}
% V_{angle} = k_\theta (\theta - \theta_{eq})^2 
% \end{equation}\\
%  \hline
%  Dihedrals & \begin{equation}
% V_{torsion} = \sum_{n=1}^{n_max}{k_n (1 + cos(n\psi - \gamma))}
% \end{equation}\\
%  \hline
%  Lennard Jones    & \begin{equation}
% V_{VdW} = \frac{A}{r^{12}} - \frac{B}{r^6}
% \end{equation}\\
%  \hline
%  Electrostatics&   \begin{equation}
% V_{electrostatics} = k_0 \frac{q_i q_r}{r}
% \end{equation}\\
%  \hline
% \end{tabular}

\paragraph{Bonds. } The potential energy of the bonds depends on the overall distances between all the atoms that have an immediate connection to one another. $k_0$ is defined as the force constant and $r$ is the distance between bonded atoms. $r_{eq}$ is the equilibrium distance between the bonded atoms. Then, $V_{bonded} = k_0 (r - r_{eq})^2$.

\paragraph{Angles. }
Angles are the second-most important potential energy of the moelcule, in which torchmd calculates $\theta$ as the angle between the three bonded atoms. $k_0$ is the angular force constant, and $\theta_{eq}$ is the equilibrium angle. Then, $V_{angle} = k_\theta (\theta - \theta_{eq})^2$. 

\paragraph{Dihedrals. }
The third important is the dihedral angles, but in TorchMD, it is called torsion. $\psi$ is the dihedral angle between the four atoms. $\gamma$ is the phase offset, and $k_n$ is the amplitude of the harmonic component of periodicity $n$. Then, $V_{torsion} = \sum_{n=1}^{n_{max}}{k_n (1 + cos(n\psi - \gamma))}$.

\paragraph{Lennard Jones. }
In this paper, we consider this Lennard Jones, whereas TorchMD calls this Van der Waals energy. In previous work, such as Ham-Net \cite{li2020conformation}, they use this equation as a loss for their model, but where A and B is proportional to the output weights of their deep learning model. Then, $V_{VdW} = \frac{A}{r^{12}} - \frac{B}{r^6}$, where, $A = 4e\sigma^{12}$ and $B = 4e\sigma^2$.

\paragraph{Electrostatics. }
Electrostatics is the Coulomb's force, such that $k_e = \frac{1}{4\pi e_0}$, and $q_i$ and $q_r$ are the charges of two atoms. Then, $V_{electrostatics} = k_0 \frac{q_i q_r}{r}$, where, $A = 4e\sigma^12$ and $B = 4e\sigma^2$.

\paragraph{Total Energy. }

TorchMD analytically calculates all the previous potential energies individually, and outputs a dictionary in which we can add all the potential energy to become the total potential energy for a given structure. We also added an external potential energy (End2End distance) that adds a force bias (given a timestep t) that we want the generated trajectory to follow. Overall, the total potential energy is defined as $V_{total} = \sum^{n_{bonds}}{V_{bonded}} + \sum^{n_{angles}}{V_{angles}} +
    \sum^{n_{torsion}}{V_{torsion}} \\ + \sum^{n_{atoms}}_{i}\sum^{n_{atoms}}_{j < i}{(V_{VdW} + V_{electrostatics})} + V_{End2End}(t)$.

% \begin{equation}
%     \begin{aligned}
%     V_{total} = \sum^{n_{bonds}}{V_{bonded}} + \sum^{n_{angles}}{V_{angles}} +
%     \sum^{n_{torsion}}{V_{torsion}} \\ + \sum^{n_{atoms}}_{i}\sum^{n_{atoms}}_{j < i}{(V_{VdW} + V_{electrostatics})} + V_{End2End}(t)
%     \end{aligned}
% \end{equation}

% $V_{ext}$ is the external forces applied to the molecule and the subsequent atoms; however in this paper, we replace $V_{ext}$ as End-to-End distance, which is a collective variable that we would want the model to learn, such that we can force the model in certain direction.

\section{Experiments and Results}

% All of our experiments are presented in this Github link: \url{https://github.com/johncava/Molecular_Dynamics}. 

Our experimental implementation is available in Pytorch \cite{paszke2017automatic} and TorchMD, which was executed on a computing cluster with NVIDIA Tesla K80s. 

\paragraph{Data and Data Augmentation}

\begin{figure*}
 \centering
 \includegraphics[width=0.8\textwidth]{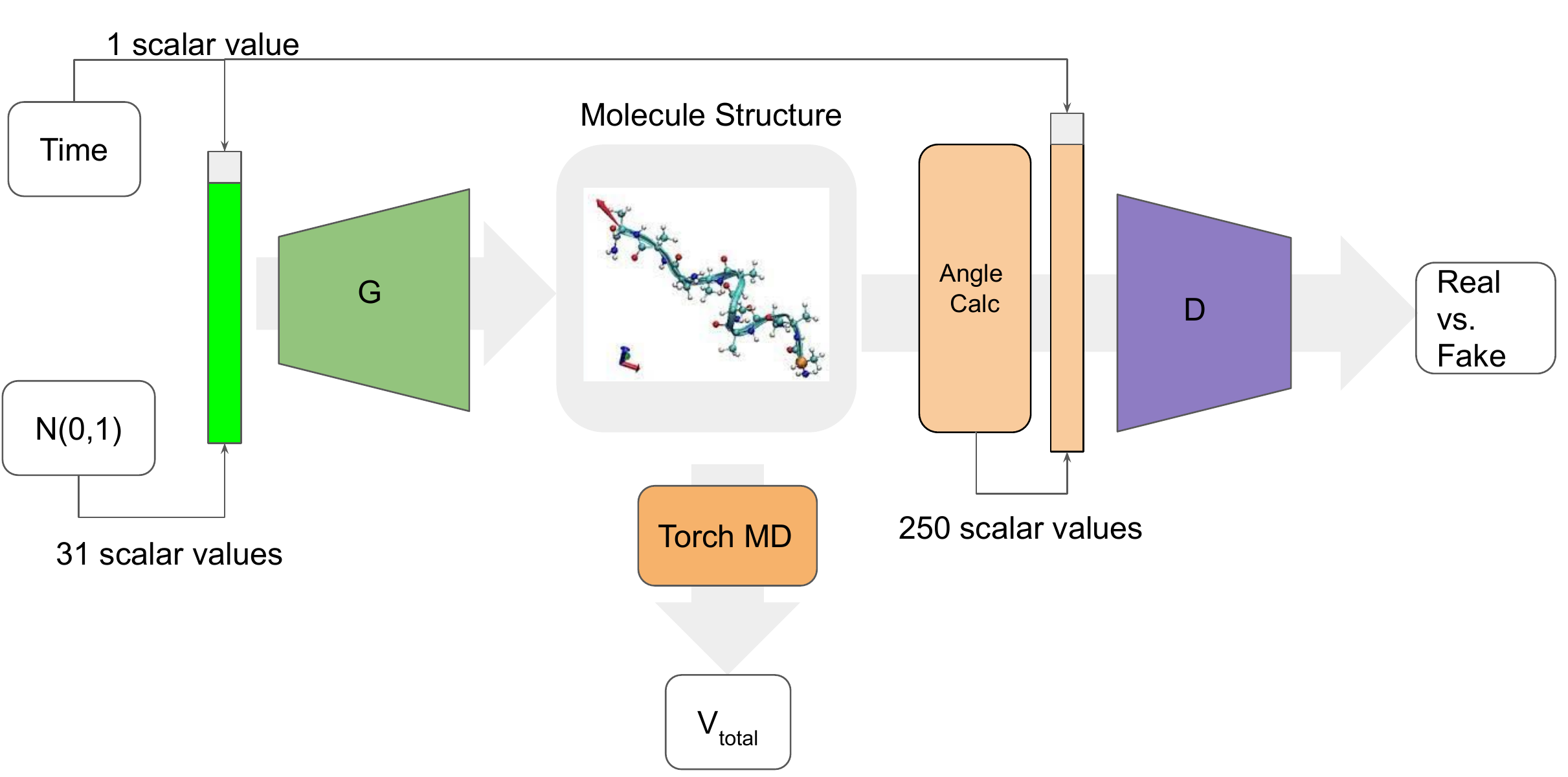}
\caption{cGAN Model. A Generator that takes in a time step and random vector, which outputs a frame that is then used as an input to the Discriminator and the calculation of the potential energy.}
\vspace{-0.2in}
\end{figure*}

%\begingroup
% \setlength{\intextsep}{0pt}%
% \setlength{\columnsep}{5pt}%
% \begin{wrapfigure}{r}{0.6\textwidth}
% %\begin{figure}[h]
% \centering
% \includegraphics[width=0.58\textwidth]{Model.pdf}
% \caption{cGAN Model. A Generator that takes in a time step and random vector, which outputs a frame that is then used as an input to the Discriminator and the calculation of the potential energy.}
% \end{wrapfigure}
% %\endgroup

For our experiments, we utilize the data for simulating a 10 deca-alanine molecule given a bias force that would stretch it. For these experiments, we conducted an augmentation such that it is randomly rotated and all the trajectories are aligned on the center of mass of the molecule. We did this in order to make sure that the generator does not over-fit on specific positions and translations.

\paragraph{Conditional GAN with Geometric Invariances.}

We take inspiration from MolGAN \cite{de2018molgan}, cGANs \cite{DBLP:journals/corr/MirzaO14}, and InvNet \cite{Joshi_Cho_Shah_Pokuri_Sarkar_Ganapathysubramanian_Hegde_2020} in order to generate structures that are conditioned on time. The loss function is augmented by the potential energy computed through TorchMD. Our model is described in Figure 1. The generator is a feed forward neural network that takes in a time step, and a normal vector of size 31. This then generates the structure which would be 40 atoms (in backbone prediction) and 104 atoms (in full system). We also add a constraint by computing the potential energy from the generated structures in order to train a better generator. 
The generator is a feedforward neural network: [Linear(32,50), Linear(50,100), Linear(100,312)]. The discriminator takes 1 time-step + 250 torsion-angles computed from the 312 sized output structure. Thus, the discriminator is structured as another feedforward neural network: [Linear(251,10), Linear(10,1)].  For the optimizer, we used Adam \cite{kingma2014adam} for optimizing the generator and discriminator (learning rate = $1\times 10^{-3}$). The number of epochs for GAN training is 10.

\begin{figure}[ht]
\centering
\includegraphics[scale=0.40]{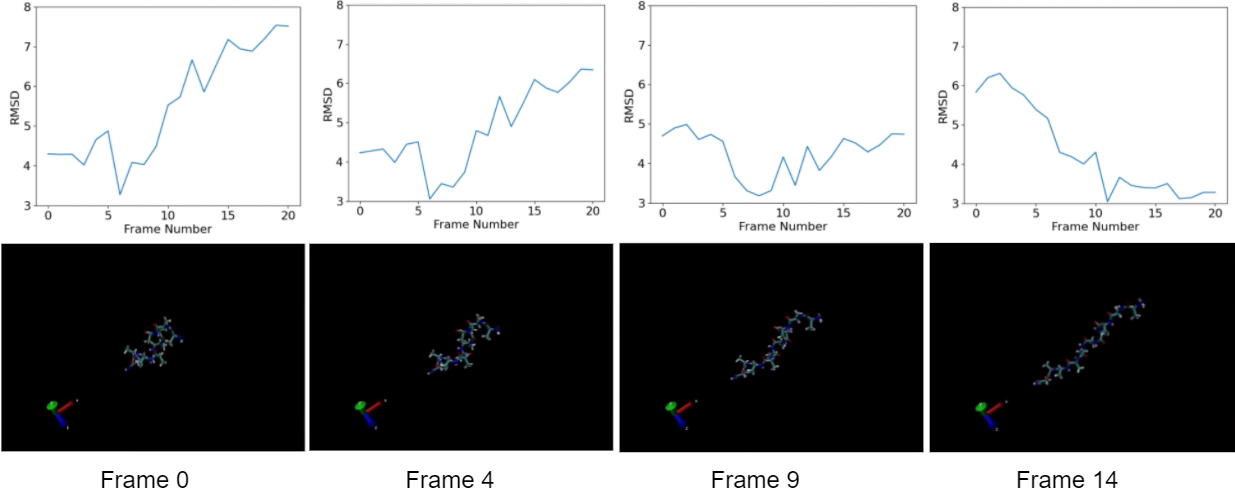}
\caption{Four generated frames trained by cGAN alongside the RMSD plots with respect to the ground truth trajectory.}
\end{figure}

\paragraph{Loss Function.}
The loss function we use is a standard cGAN loss, augmented with the total potential energy term described in section \ref{Sec:EnergyTerm}. The loss function is then,

\begin{equation}
   \min_G \max_D L(D,G) + \lambda V_{total},
\end{equation}

where, $L(D,G)$ is the standard cGAN loss function, conditioned on time-step $t$, given by $\mathbb{E}_{x \sim p_{data}(x)}[log D(x|t)] + \mathbb{E}_{x \sim p_{z}(z)}[log[ 1 - D(G(z|t))]$. Due to the additional term $V_{total}$, which actually depends on the generator parameters, a total derivative of the loss function is difficult. Motivated by the three-way update
rule that was used in InvNet \cite{Joshi_Cho_Shah_Pokuri_Sarkar_Ganapathysubramanian_Hegde_2020}, we find that a similar strategy works well even in our case. That is, a GAN-like update of $G$ via gradient steps
of $L(D,G)$ keeping $D$ fixed; a GAN-like update of $D$ via gradient steps of $L(D,G)$ keeping $G$ fixed; and an update of $G$
via gradient steps of $V_{total}$.

\paragraph{Pre-Training the GANs with Annealing.} In order to narrow down the search space, we implement pretraining of the generator before we do the adversarial training. In addition, we conduct an annealing process for the pretraining. In this process, we slowly transition from training on reconstruction (of position/angles) to potential energies (bonds, angles, dihedrals). In the reconstruction loss, we have an emphasis on torsion angles such that we emphasize the initial structure of the helix for the deca-alanine. Thus once we do the adversarial training, we then emphasize on the End2End potential energy bias force exerted on the deca-alanine. The number of epochs for pretraining is 5, and we used a learning rate of $1 \times 10^{-2}$.

\paragraph{Adversarial Training and Potential Encoding.}

In our adversarial training, we emphasize an invariant conditional component (as inspired by physical constraint GANs such as invGAN). Instead of a neural network, we use potential energies calculated from TorchMD given the generated structures. We have experimented with a hierarchy of potential energies given a certain epoch. Where we focus in order of bonds, angles, dihedrals, impropers, LJ, and electrostatics. Together, they enforce the physical constraints that make the generator learn how to make a physically realistic structure. We have also made it such that with pretraining we have the generator creating good enough structure based on the bonds, angles, and dihedrals, since LJ is sensitive to plainly wrong structures, which would produce high potential energy, and thus blow up the gradients. In addition, to avoid blowing up of gradients, we did gradient clipping on the generator such that during training the potential energy loss wouldn't blow up the gradients for the generator.

\paragraph{Results and Conclusion.}

After training the cGAN, we generated 20 time points conditioned on the maximum trajectory size of 1002 frames. In Figure 2, we present some exemplary generated time points from our cGAN model. The lowest point of each RMSD plot is the `ground truth match' to the generated frame.
We find that as the time points progress from 4 to 9 to 14, our generated models match frames 6, 8 and 11 of the ground truth data (downsampled from 1002 to 20 frames). Based on these frames, we can conclude that the model is able to find the minimial action potential pathway from a favorable conformational state to a less conformational state. We have tried training the cGAN without the $V_{total}$, and we were not able to generate physically realistic structures.

However, there are still limitations in which the starting structure at time point 0 is not helical, which indicates that there needs to be better ways of representing structure as adding a discriminator and potential energy loss constraint aren't enough to enforce that initial structure. Future work could be considering representing the input as a ball and stick model and treat changes as `actions' in a reinforcement learning framework, or also representing the input as a spatio-temporal graph. 

\bibliographystyle{plain} % We choose the "plain" reference style
\bibliography{ref} % Entries are in the refs.bib file
% \small{\bibliographystyle{ieee}
% \bibliography{ref}
% }
\end{document}